\documentclass{agujournal}

\journalname{Space Weather}

\begin{document}

%
%


\title{Flare forecasting at the Met Office Space Weather Operations Centre}

%
%




\authors{S. A. Murray\affil{1}, S. Bingham\affil{2}, M. Sharpe\affil{2}, and D. R. Jackson\affil{2}}


\affiliation{1}{Trinity College Dublin, Ireland}
\affiliation{2}{Met Office, United Kingdom}





\correspondingauthor{S. A. Murray}{sophie.murray@tcd.ie}




\begin{keypoints}
\item Flare forecasts produced daily at the Met Office Space Weather Operational Centre
\item Forecasts since 2014 verified using numerical weather prediction methods
\item Clear benefit to human `influence' on issued flare forecasts
\item Forecast skill shown to decrease for longer forecast lead times
\item Real time verification has been implemented for forecaster use
\end{keypoints}

%
%


\begin{abstract}
The Met Office Space Weather Operations Centre produces 24/7/365 space weather guidance, alerts, and forecasts to a wide range of government and commercial end users across the United Kingdom. Solar flare forecasts are one of its products, which are issued multiple times a day in two forms; forecasts for each active region on the solar disk over the next 24 hours, and full-disk forecasts for the next four days. Here the forecasting process is described in detail, as well as first verification of archived forecasts using methods commonly used in operational weather prediction. Real-time verification available for operational flare forecasting use is also described. The influence of human forecaster is highlighted, with human-edited forecasts outperforming original model results, and forecasting skill decreasing over longer forecast lead times.
\end{abstract}

%
%

%


%
%
%
%

\section{Introduction}
\label{intro}

Space weather forecasting has developed rapidly in recent years, with the threat of a severe space weather event increasing in importance as society becomes ever more dependent on technology. Space weather service providers worldwide have developed monitoring systems for solar events of particular interest to space weather, namely solar flares, coronal mass ejections (CMEs), and solar energetic particle events (SEPs). Solar flares impact near-Earth space within minutes, while SEPs can take tens of minutes, and CMEs days, to reach Earth. Forewarning of solar flares is thus particularly important for operational space weather services. 

Whilst CME and SEP forecasting is still in its infancy \citep{zheng13, luhmann15}, much research has already been undertaken in the field of solar flare forecasting. Early work focused on statistical methods based on historical flaring rates \citep{mcintosh90, gallagher02, leka07, wheatland05}, however more complex methods have been developed in recent years \citep{barnes07, georgoulis07}, and sophisticated computational techniques such as machine learning \citep{ahmed13, bobra15} have become popular. \citet{barnes16} systematically compared many of these various types of methods, and found no one method clearly outperformed all others. More significantly, \citeauthor{barnes16} also found no method proved substantially better than climatological forecasts. 

With more work needed by the solar physics community to improve the accuracy of current flare forecasting methods, most operational space weather forecasting centres worldwide still rely on human forecasters to issue alerts, warnings, and forecasts. NOAA's Space Weather Prediction Centre (SWPC) has been operating since 1946, providing daily forecasts for Geostationary Operational Environmental Satellite (GOES) soft X-ray M- and X- class flares for the next three days. Since SWPC's inception many other operational centres have been set up worldwide with forecasters providing daily guidance, for example the Royal Observatory of Belgium's Solar Influences Data Analysis Centre (SIDC), and the South African National Space Agency. Others have moved towards an automated service, such as the Korean Space Weather Center's Automated Solar Synoptic Analyser (ASSA; \hbox{http://spaceweather.rra.go.kr/models/assa}), however they are less prevalent than human-issued forecasts in operational space weather centres. The International Space Environment Service webpage (\hbox{http://www.spaceweather.org}) shows many current space weather centres worldwide.

A severe space weather event was added to the United Kingdom's (UK) National Risk Register of Civil Emergencies in 2011. The Met Office, the national meteorological service for the UK, were given ownership of that risk in 2013 and set up the Met Office Space Weather Operations Centre (MOSWOC) to provide space weather alerts, warnings, and guidance to the UK government and general public. The centre was officially opened in 2014 October, although 24/7 operational services commenced in 2014 April. MOSWOC provides flare forecasts to users multiple times daily as part of their space weather service. This paper will outline the method behind these forecasts (Section~\ref{method}), the first verification results of archived forecasts (Section~\ref{validation}), as well as outlining future plans for forecast development (Section~\ref{discussion}).


\section{MOSWOC Flare Forecasts}
\label{method}

Before calculating any flare probabilities, a MOSWOC forecaster first undertakes a thorough analysis of current solar conditions using images from Solar Dynamics Observatory's Heliospheric Magnetic Imager \citep{scherrer12}. In particular, magnetograms (e.g., \hbox{http://sdo.gsfc.nasa.gov/assets/img/latest/latest\_4096\_HMIBC.jpg}) and intensitygrams (e.g., \hbox{http://sdo.gsfc.nasa.gov/assets/img/latest/latest\_4096\_HMIIF.jpg}) are overlayed to compare the magnetic structure of active regions (ARs) to the visible sunspots. Each SWPC-numbered region on disk, as well as any other region of interest to the forecaster (often newly emerged regions that have not yet been numbered by SWPC) are analysed. The forecaster manually assigns Modified Mount Wilson \citep{kunzel65} and McIntosh \citep{mcintosh90} classifications to each region. The location, length, and area are also manually determined using internal software. The identified ARs are added to solar synoptic maps created regularly by the forecaster, which also contain coronal hole, filament, and polarity inversion line information. 

Flare probabilities are then calculated based on historical flare rates for each McIntosh class. MOSWOC uses a database containing GOES X-ray flare and McIntosh classifications for this purpose from various sources:
\begin{itemize}
\item{Kildhahl data from 1969 to 1976 supplied by \citet{bloomfield12}.}
\item{SWPC data from 1988 to 1996 supplied by \citet{bloomfield12}.}
\item{Data from 1996 to 2011 supplied by the ASSA system.}

\end{itemize}
This database is used to calculate an average daily flare rate, $\mu$, for each McIntosh classification \citep[see][for more details]{bloomfield12}. The MOSWOC forecaster then calculates the flare probabilities for M- and X- class flares using the Poisson statistics technique of \citet{gallagher02}, where the probability of observing one or more flares in a 24-hour period is $1 - e^{-\mu}$. The resulting probabilities for each classified region are combined to give a full-disk percentage probability,
 \[  Total~\% = 100(1 - \prod\limits_{n=1}^{N}(1 - \frac{AR_n~\%}{100}))\]
where $N$ is the total number of ARs identified. These percentage probabilities are used as a basis for the flare forecasts issued to end-users by MOSWOC. It is worth noting that the same method is used for automatic flare forecasts issued at SolarMonitor.org (\hbox{http://solarmonitor.org/forecast.php}), except only the \citet{bloomfield12} data is used to calculate average flare rates.


\subsection{Current Products}

\begin{figure}
\centering
\noindent\includegraphics[width=\textwidth]{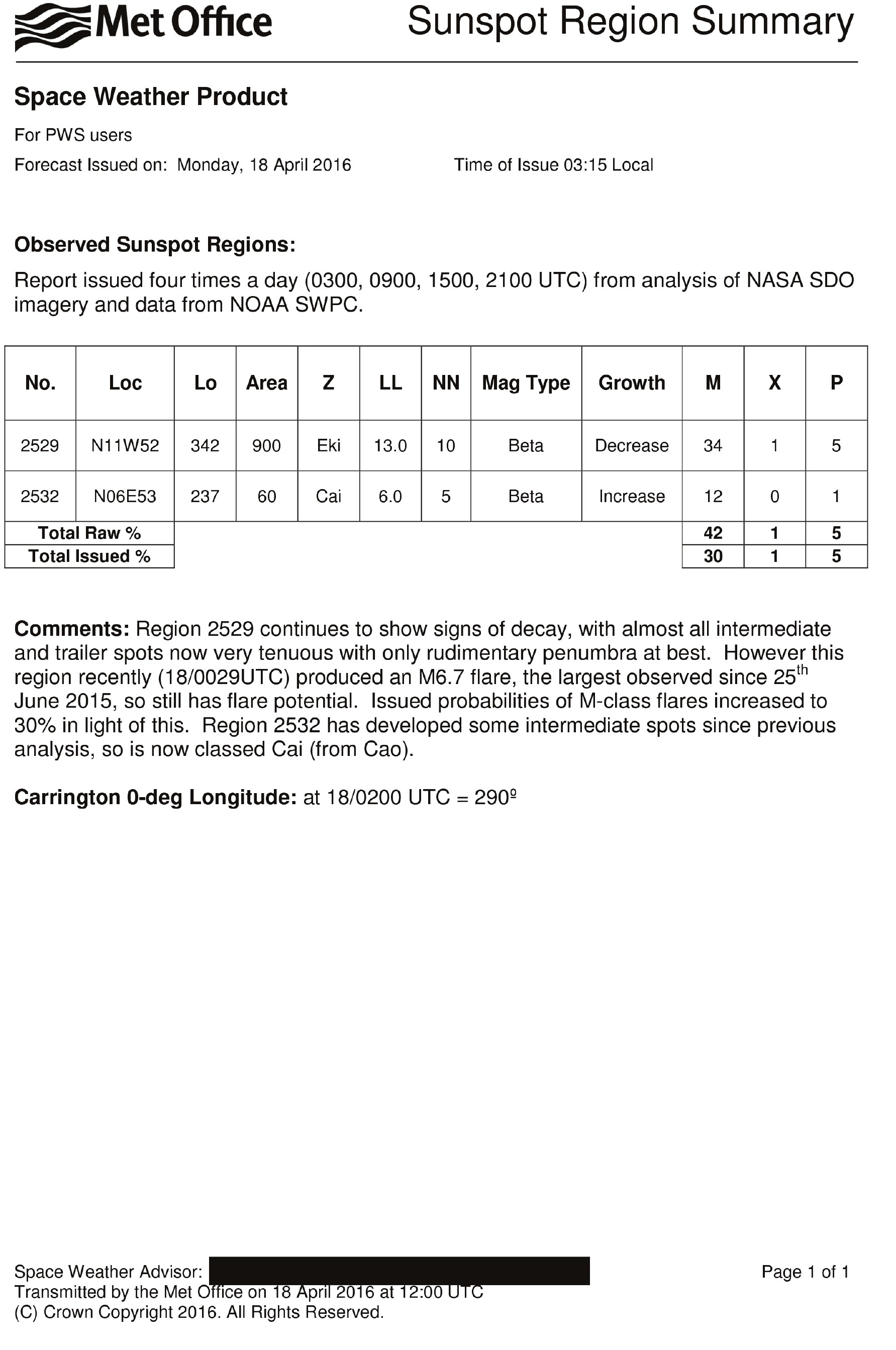}
\caption{Sunspot Region Summary issued by MOSWOC on 2016 April 18 valid from 03:00 UTC. The previously issued total issued M and X forecasts were 20\% and 1\% respectively (valid from 2016 April 17 21:00 UTC).}
\label{method:srs_example}
\end{figure}

The method above is carried out by MOSWOC forecasters every six hours, since Sunspot Region Summaries (MOSWOC-SRS) are issued every six hours to users via email. The MOSWOC-SRS is similar in format to the SWPC Solar Region Summaries (NOAA-SRS), containing sunspot properties as well as M- and X- class flare forecasts (see Figure~\ref{method:srs_example} for an example). The properties listed within the MOSWOC-SRS are described in Table~\ref{method:srs_table}. The values in the `M' and `X' columns for each active region listed are those percentage probabilities calculated from the Poisson method described above. Each probability forecast is valid from the issue time (3, 9, 15, and 21 UTC) for the immediate next time interval of 24 hours. The full-disk probability calculated from this method is also listed in the `Total Raw \%' row in the MOSWOC-SRS. It is worth noting however that the official percentage probabilities issued by MOSWOC are listed in the `Total Issued \%' row, which is based on human judgement. The forecaster has the option here to manually edit the model results if they feel it is necessary based on their own experience after examining all available data. For example if the X-ray flux is low, sunspots are declining in size, etc, then the model value might be lowered before issuing.

\begin{table}
\caption{Description of parameters provided by the MOSWOC Sunspot Region Summary table.} 
\centering 
\begin{tabular}{l l} 
\vspace{0.1cm} \\
\hline
No 				& Region number determined by SWPC (labelled xxx, yyy, etc if not numbered).\\
Loc 			& Sunspot group latitude and Stonyhurst longitude on the visible disc in heliographic degrees. 		\\
Lo 				& Carrington longitude location of the region.								\\
Area 			& Total estimated area of the group in millionths of the solar hemisphere.	\\
Z 				& McIntosh classification of the group.								\\
LL 				& Longitudinal extent of the group in heliographic degrees.					\\
NN				& Total number of individual sunspots visible in the group.					\\
Mag Type		& Modified Mount Wilson classification of the group.						\\
Growth			& Trend of region development (`Nil', `Decrease' or `Increase').			\\
M 				& Percentage risk of M-class flares.										\\
X				& Percentage risk of X-class flares.										\\
P				& Percentage risk of proton storms.											\\
Total Raw		& Accumulative percentages of all regions.									\\
Total Issued	& As above but human-edited if necessary.									\\
\hline
\end{tabular} 
\label{method:srs_table} 
\end{table}

\begin{sidewaysfigure}
\centering
\noindent\includegraphics[width=\textwidth]{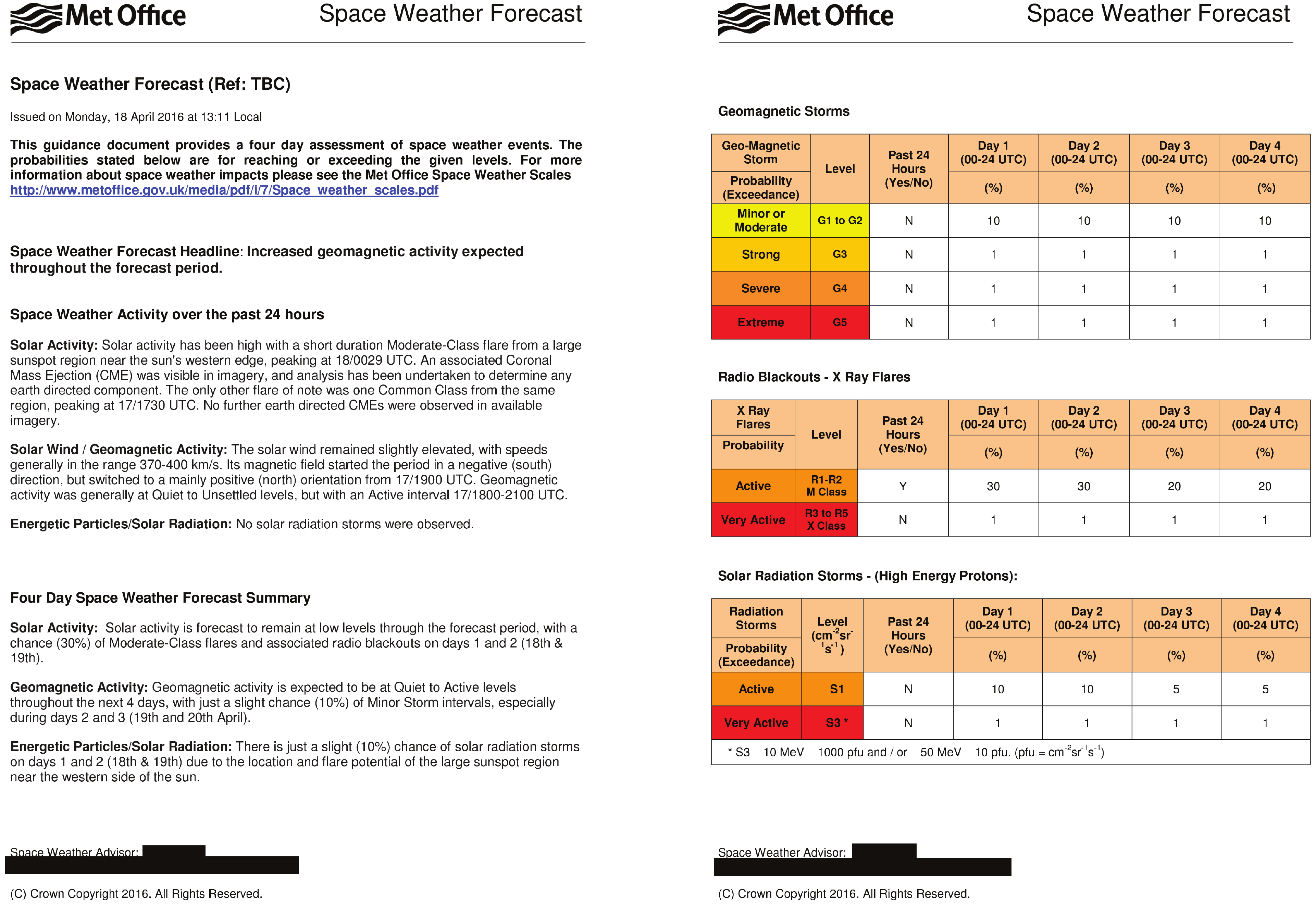}
\caption{Space Weather Forecast issued by MOSWOC on 2016 April 18. The document was issued at 13:11 local time, with forecasts being valid from 12:00 UTC. Note that the reference in the main title is for internal purposes only. The forecaster purports to GOES C- and M-class flares when referring to `common class' and `moderate class' respectively in the text.}
\label{method:guidance_example}
\end{sidewaysfigure}

The total issued forecasts in the MOSWOC-SRS are a basis to the flare forecasts issued by MOSWOC in their Space Weather Guidance documents. These are available in two forms - a technical document for those with a strong understanding of space weather and solar physics, and a simpler document for everyday users. The technical Space Weather Forecast document provides a four day assessment of space weather events, including geomagnetic storms, radio blackouts from X-ray flares, and solar radiation storms in the form of high energy protons ($\geq 10$~MeV and $\geq 100$~MeV) and high energy electrons ($\geq 2$~MeV). It is issued as a pdf document via email twice daily at 00~UTC and 12~UTC. The document includes observations of events over the past 24 hours, together with the forecast for the coming 24-hour period, and a further 3-day outlook. Scientific explanations accompany the forecasts for the more advanced user, and forecasters can also include any figures that may be relevant to their descriptions, such as the latest solar synoptic map, model output, or spacecraft data. See Figure~\ref{method:guidance_example} for an example of the simpler document, which does not include the more complex textual descriptions and figures, and the high energy electron forecasts are also removed due to end-user requirements. The simpler document is available via email subscription twice daily, and the information is also available online for registered users at the Met Office public web pages (\hbox{http://www.metoffice.gov.uk/space-weather}), including sector-specific webpages.

Note that the flare forecasts in Figure~\ref{method:guidance_example} are titled `Radio Blackouts', since this is what end-users for this product are concerned with, however the values listed are X-ray flare percentage probabilities. To remove any confusion for the reader, these forecasts will henceforth be referred to as `Issued Flare Forecasts' (IFF). For these guidance documents the Day 1 M- and X- class IFF are obtained from human-edited full-disk forecasts issued in the most recent MOSWOC-SRS. As the MOSWOC-SRS would be calculated a few hours before this forecast is issued, the forecasters will generally manually edit this value to reflect the recent evolution of these ARs. The Day 2, Day 3, and Day 4 forecasts are then determined by forecaster experience based on how the ARs are evolving, and what ARs may be leaving or returning to the solar disk in the next few days. It is worth noting that the midnight forecast for each day is valid for 24~hours only, i.e., the next 24~hours, 24 - 48~hours, 48 - 72~hours, and 72 - 96~hours. Forecasts for both MOSWOC-SRS and IFF are given for M 1.0 - 9.9 (above M- but below X- class), and for X 1.0 and above.


\section{Forecast Validation}
\label{validation}

Historically the solar physics community has used categorical verification techniques to validate new forecast methods. This entails deciding a threshold at which the probabilistic values become a `yes/no' forecast and then calculating metrics such as the Heidke Skill Score and True Skill Score \citep[][]{barnes08, crown12, bloomfield12}. More recently however the community has looked to operational meteorological verification techniques more suitable for probabilistic forecasting, increasingly presenting reliability diagrams and relative operating characteristic (ROC) curves alongside these traditional skill scores \citep[][]{guerra15, barnes16, cui16}. See the World Weather Research Programme/Working Group on Numerical Experimentation (WWRP/WGNE) Joint Working Group on Forecast Verification for a comprehensive description of current methods (\hbox{http://www.cawcr.gov.au/projects/verification/}).

Reliability diagrams measure how closely the forecast probabilities of an event correspond to the actual chance of observing the event. The reliability diagram is conditioned on the forecasts, plotting frequency of the observations against the forecast probability to give information on the real meaning of the forecast. It is thus a good partner to the ROC curve measuring forecast discrimination, which is conditioned on the observations. ROC curves provide information on the hit rates and false alarm rates that can be expected from use of different probability thresholds to trigger advisory action. ROC curves can be used to select the trigger threshold for an event that provides the best trade-off between hit rate and false alarm rate for a particular type of decision. Here we present these results for both the MOSWOC-SRS and IFF. 

The MOSWOC-SRS have been archived since July 2015, therefore a year's worth of these forecasts were analysed between 2015 - 2016 July 15. All the MOSWOC IFF have been archived since January 2014, thus approximately 31 months of these forecasts were analysed between 2014 January 1 - 2016 July 15. In particular, each of the four MOSWOC-SRS and two IFFs issued daily were used, all being valid for 24 hours from their time of issue. All M- and X- class full-disk flare forecasts were compared to the GOES X-ray flare events recorded in the SWPC Solar and Geophysical Event Reports (ftp://ftp.swpc.noaa.gov/pub/indices/events/) during these time periods. There were unfortunately not enough X-class flares observed during the forecast periods for any meaningful analysis (0 X-class flares observed for MOSWOC-SRS and 18 for IFF), thus results highlighted in this paper are only for the M-class flare forecasts (57 M-class flares observed for MOSWOC-SRS and 338 for IFF).


\subsection{MOSWOC Sunspot Region Summary results}

\begin{figure}
\centering
\noindent\includegraphics[width=\textwidth]{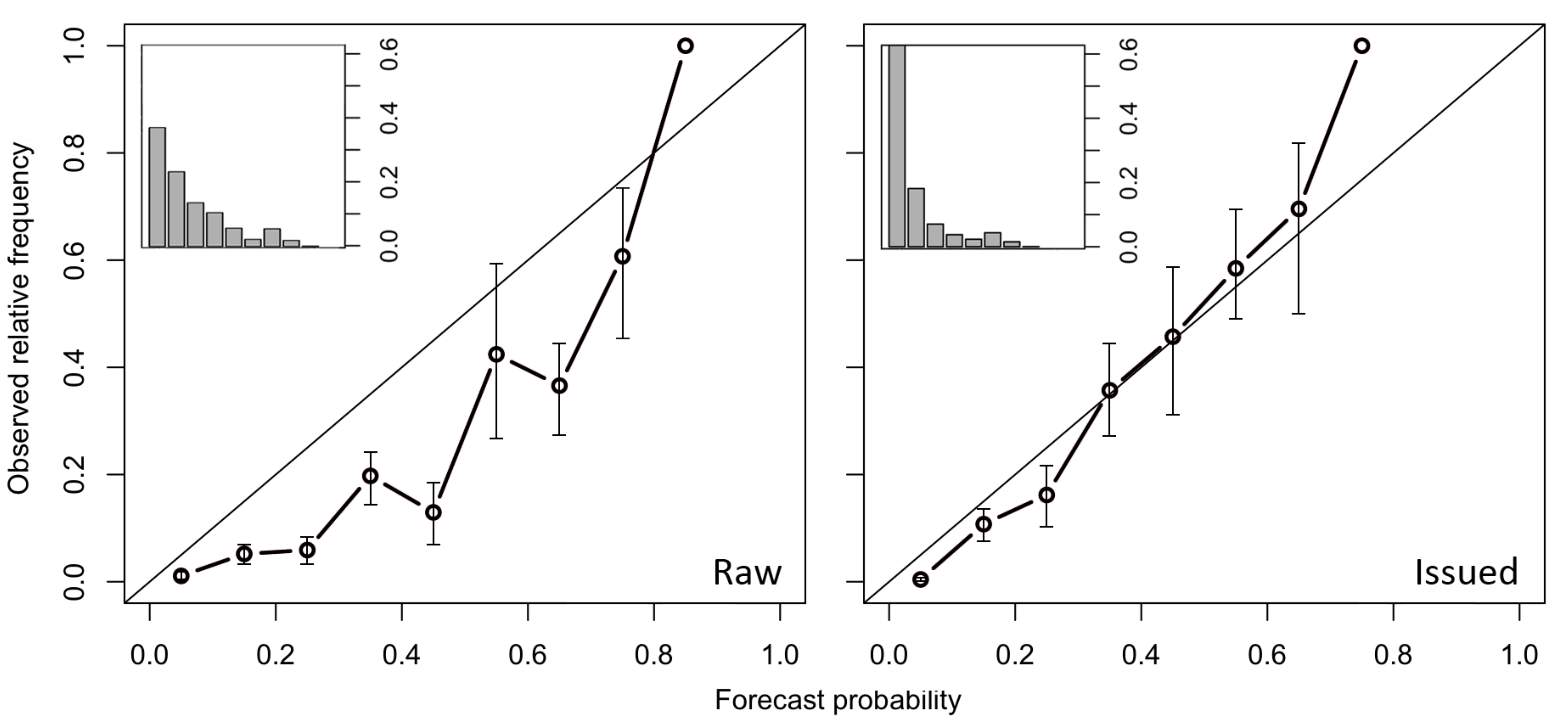}
\caption{Reliability diagrams for MOSWOC-SRS raw (left) and issued (right) forecasts (for the next 24 hours) for a one year period beginning 2015 July 15. The subplot shows the distribution of forecasted probabilities for the time period studied.}
\label{srs_m_reliability}
\end{figure}

Figure~\ref{srs_m_reliability} shows the reliability diagrams for the MOSWOC-SRS full-disk raw (left) and human-edited (right) forecasts. Data in the main and sub plots are binned the same conditional to the forecast probability, less bins for the human-edited forecast with the biggest forecasted probability for the time period 75\% compared to 81\% for the raw forecast. For perfect reliability the forecast probability and the frequency of occurrence should be equal, and the plotted points should lie on the diagonal line. The raw probability points mainly lie below the diagonal line, highlighting a tendency to over-forecast. It is clear that the human influence on issued probabilities has resulted in improvement upon the model results, with points lying closer to the diagonal. The distributions in the subplots of Figure~\ref{srs_m_reliability} highlight that the forecasters tend to decrease the probability values, leading to less over-forecasting in general. The human-edited forecasts do not improve the case of the highest probability bins however, with both reliability diagrams showing under-forecasting for M-class flares in the next 24 hours. This may be related to the lack of data for higher probabilities, with very few forecasts issued at higher probabilities (the rarest in the data set).

\begin{figure}
\centering
\noindent\includegraphics[width=0.75\textwidth]{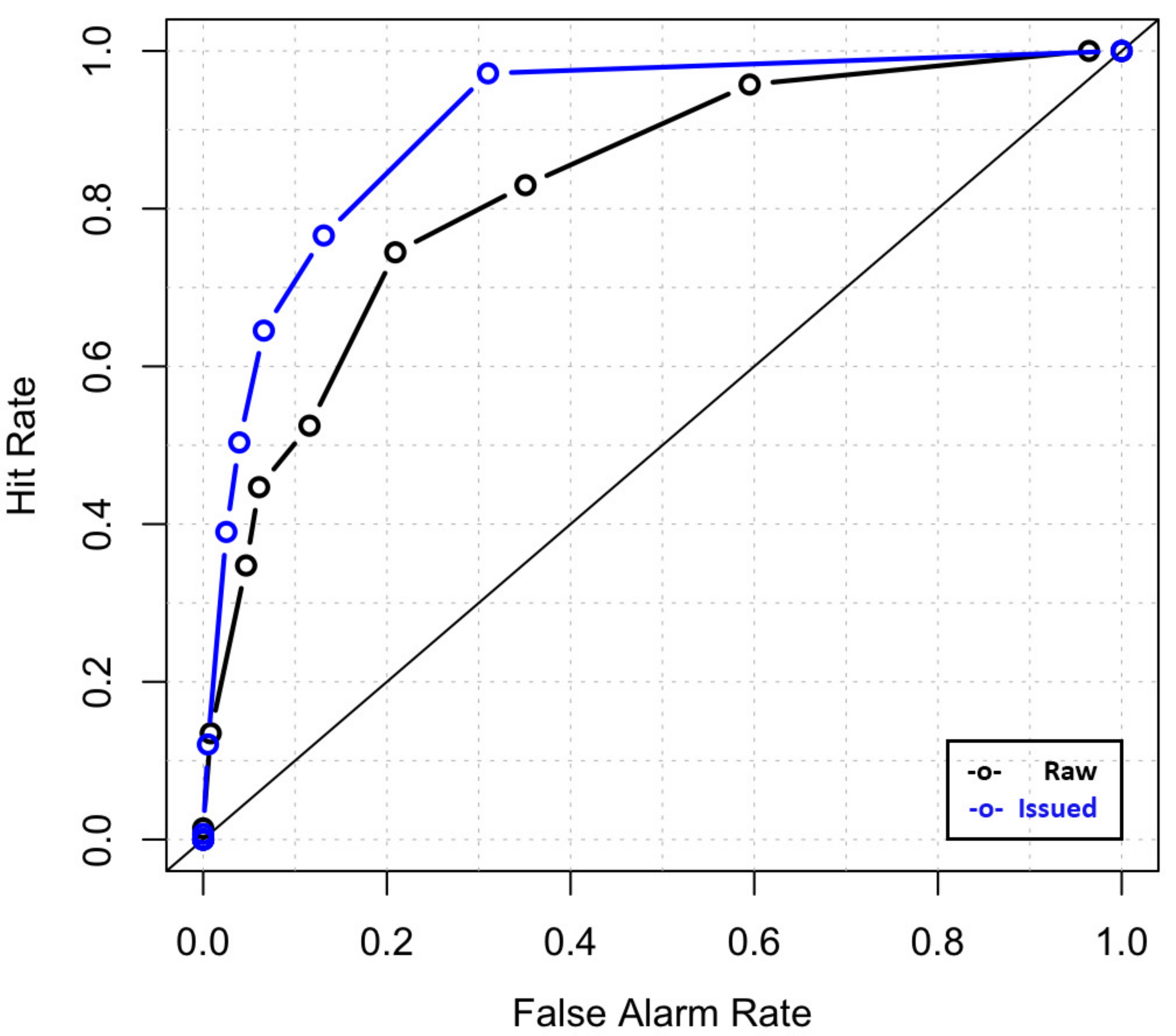}
\caption{Relative Operating Characteristic curves for MOSWOC-SRS raw (black) and issued (blue) forecasts (for the next 24 hours) for a one year period beginning 2015 July 15. The optimal thresholds for the raw and issued forecasts are 0.265 and 0.175, respectively. Note that the curves in this figure show the results of thresholds taken every 0.1 between 0 and 1.}
\label{srs_m_roc}
\end{figure}

ROC curves for the MOSWOC-SRS raw and issued forecasts are shown in Figure~\ref{srs_m_roc}, including values for the optimal threshold that maximises both sensitivity and specificity \citep[Youden's index; ][]{youden50}. A skillful forecast system will achieve hit rates that exceed the false alarm rate, thus the closer the curve is to the top left corner of the plot, the more skillful the forecast. For the case of the MOSWOC-SRS curves, it is clear that the issued forecast (right) shows more skill than the raw forecast (left). Thus, similar to the reliability diagram results, the human-influenced forecasts have improved upon the Poisson method output.

The verification results for the MOSWOC-SRS are summarised in Table~\ref{table:srs}. The area under a ROC curve provides a useful summary statistic of the discriminatory ability; an area of 1.0 represents a perfect test, and an area of 0.5 represents a worthless test. For the raw vs human-edited forecasts we see a smaller ROC area for the Poisson output, indicating that a human has more skill than the model at correctly distinguishing X-ray flare events from non-events. Brier score is also listed, which measures the mean squared probability error, and gives an idea of the magnitude of the probability forecast errors (perfect score being 0). This can be partitioned into reliability (closer to 0 indicates better reliability), resolution (closer to 1 indicates better resolution), and uncertainty, which are also listed \citep{murphy73}. As explained by the WWRP/WGNE group, the reliability measures the average agreement between forecast probability and mean observed frequency; the resolution shows the ability of the forecast to resolve the set of sample events into subsets with characteristically different outcomes (i.e., the forecast has resolution if it can successfully separate one type of outcome from another); the uncertainty shows the variability of the observations (with greater uncertainty, owing to small event rates, making the forecast more difficult). These other verification metrics also confirm the human-edited forecasts outperform the model output, although for some measures only a small difference is found.

\begin{table}
\caption{Verification statistics for Sunspot Region Summary M-class full-disk forecasts.} 
\centering 
\begin{tabular}{l c c c c c c c c c} 
\vspace{0.1cm} \\
\hline\hline
Forecast  	& Total no. 	& No. of 	& No. of 		& ROC 	& Brier  	& Reliability 	& Resolution 	& Uncertainty	\\
type		& of records	& events 	& non-events 	& area	& score 	&  				& 				&  				\\
\hline
Raw				& 1489 					& 141 				& 1348				& 0.83 		& 0.090 		&  0.021 		& 0.017 		& 0.086 		\\ 
Issued			& 1489					& 141				& 1348				& 0.92 		& 0.060 		&  0.002 		& 0.028 		& 0.086 		\\ 
\hline
\hline
\end{tabular} 
\label{table:srs} 
\end{table}


\subsection{Issued Flare Forecast results}

\begin{figure}
\centering
\noindent\includegraphics[width=\textwidth]{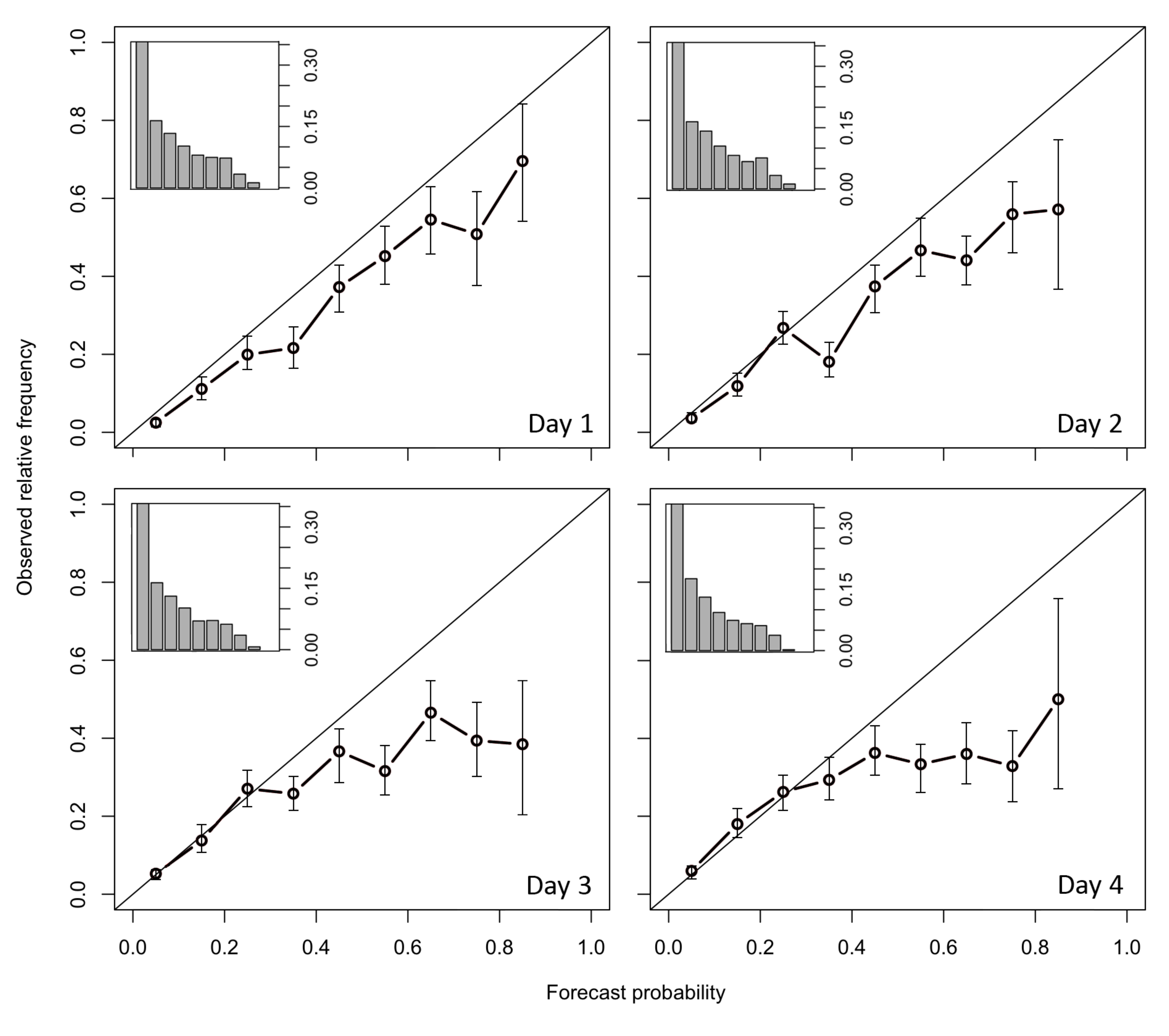}
\caption{Reliability diagrams for full-disk Day 1 (upper left), Day 2 (upper right), Day 3 (lower left), and Day 4 (lower right) Issued Flare Forecasts between 2014 January 1 - 2016 July 15. The subplot shows the distribution of forecasted probabilities for the time period studied.}
\label{tot_m_reliability}
\end{figure}

\begin{figure}
\centering
\noindent\includegraphics[width=0.75\textwidth]{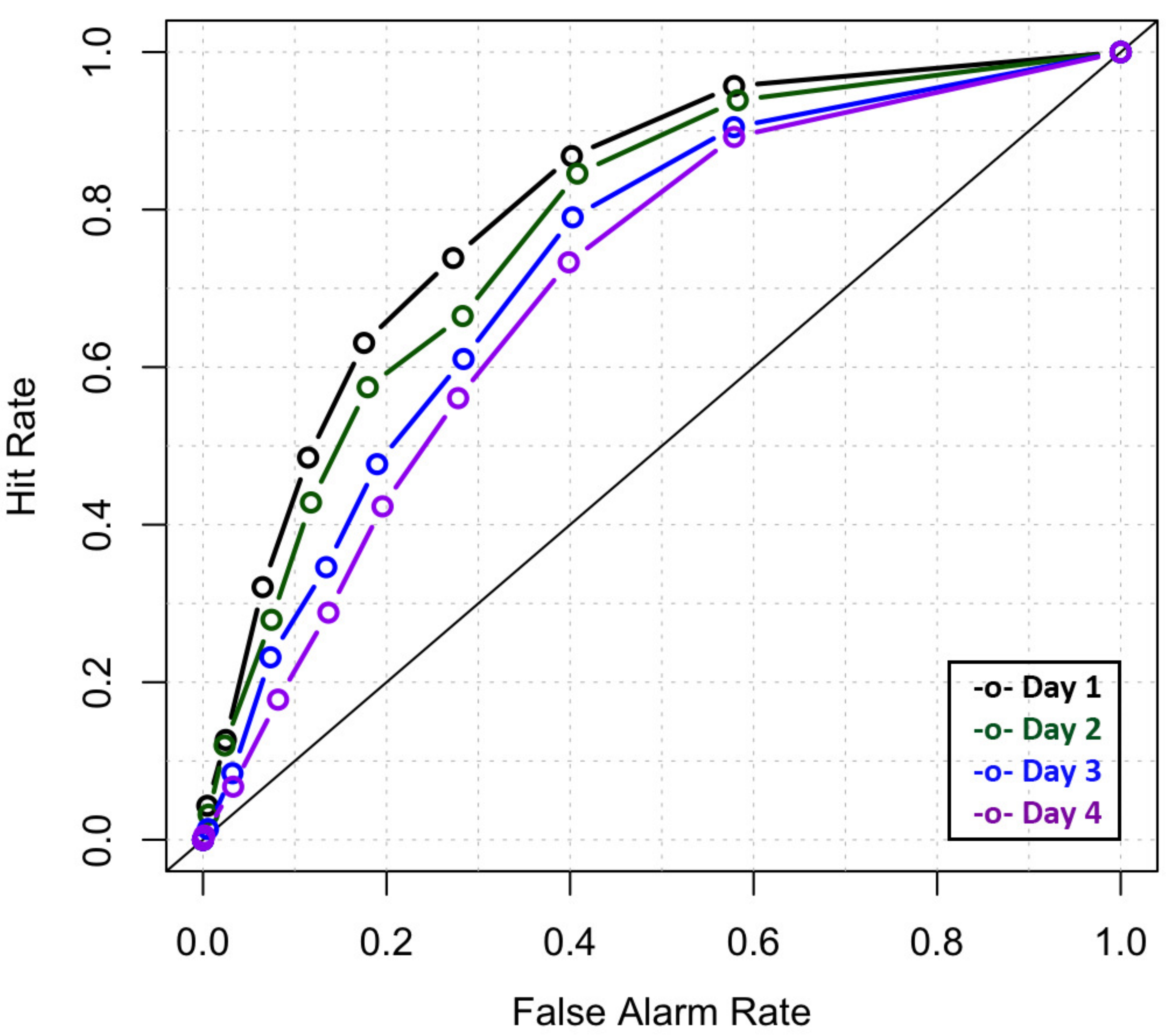}
\caption{Relative Operating Characteristic curves for full-disk Day 1 (black), Day 2 (green), Day 3 (blue), and Day 4 (purple) Issued Flare Forecasts between 2014 January 1 - 2016 July 10. The optimal thresholds for the Day 1-4 forecasts are 0.275, 0.225, 0.225, and 0.175, respectively. Note that the curves in this figure show the results of thresholds taken every 0.1 between 0 and 1.}
\label{tot_m_roc}
\end{figure}

The same analysis has been undertaken on the IFF, with reliability diagrams for the four-day forecasts shown in Figure~\ref{tot_m_reliability}. In general forecasts issued with probabilities greater than $\sim$30\% appear to over-predict flares. It is clear that the Day 1 forecast is most reliable. as this tendency becomes more pronounced for forecasts on later days. The flatter the curve in the reliability diagram, the less resolution the forecast has, thus by Day 4 the forecasts are tending toward climatology. A similar picture is presented by the ROC plots in Figure~\ref{tot_m_roc} - the best results are found on Day 1, with the ROC curves tending further toward the `no skill' diagonal as the days progress.

Table~\ref{table:iff} summarises the IFF verification scores, confirming the trends shown in Figures~\ref{tot_m_reliability} and \ref{tot_m_roc} that the forecasting skill decreases from Days 1 - 4. Values in this table indicate less skill in general compared to the MOSWOC-SRS results of Table~\ref{table:srs}, and it is worth highlighting here the differences between the data sets, with more forecasts available for the IFF analysis over a longer time period (now including all of 2014, 2015, and early 2015), and the forecasts issued only twice rather than four times a day at different times. The same verification analysis for the IFF data time period was repeated for the SWPC M-class forecasts obtained from the `Report and Forecast of Solar and Geophysical Activity' archive (ftp://ftp.swpc.noaa.gov/pub/forecasts/RSGA). Note that the SWPC forecasts are 3-day rather than 4-day forecasts, and are issued only once per day for flares M1.0 and larger. In fact, SWPC also provide their forecasts in a radio blackout forecast format similar to the MOSWOC guidance as shown in Figure~\ref{method:guidance_example} (\hbox{http://www.swpc.noaa.gov/products/3-day-forecast}). Table~\ref{table:iff} shows that the MOSWOC and SWPC results are very similar, particularly for the ROC areas, with MOSWOC only marginally outperforming SWPC on some days. Decreasing skill is also apparent for the SWPC forecasts from Days 1 - 3 for the data set analysed.

\begin{table}
\caption{Verification statistics for full-disk M-class Issued Flare Forecasts.} 
\centering 
\begin{tabular}{l c c c c c c c c c} 
\vspace{0.1cm} \\
\hline\hline
Forecast  	& Total no. 	& No. of 	& No. of		& ROC 	& Brier  	& Reliability 	& Resolution 	& Uncertainty	\\
period		& of records 	& events 	& non-events 	& area	& score 	&  				& 				&  				\\
\hline
MOSWOC Day 1	&	1864			& 371				& 1493					& 0.82 	& 0.133 		& 0.007 		& 0.033 		& 0.159 \\
MOSWOC Day 2	&	1864			& 376				& 1488					& 0.78 	& 0.142			& 0.009 		& 0.028 		& 0.161 \\ 
MOSWOC Day 3	&	1864			& 367				& 1497					& 0.74 	& 0.153 		& 0.014 		& 0.018 		& 0.158 \\ 
MOSWOC Day 4	&	1864			& 371				& 1493					& 0.71 	& 0.162 		& 0.016 		& 0.014 		& 0.159 \\ 
\hline
SWPC Day 1		&	923				& 183				& 740					& 0.80 	& 0.136 		& 0.009 		& 0.033 		& 0.159 \\
SWPC Day 2		&	923				& 183				& 740					& 0.78 	& 0.143			& 0.010 		& 0.026 		& 0.159 \\ 
SWPC Day 3		&	923				& 182				& 741					& 0.75 	& 0.150 		& 0.011 		& 0.019 		& 0.158 \\ 
\hline
\hline
\end{tabular} 
\label{table:iff} 
\end{table}


\subsubsection{Real-time verification}
\label{realtime}
MOSWOC forecasters have access to local web pages showing real-time verification information about their flare forecast products. Forecasters can examine the current prediction performance before issuing new probabilities. The idea behind these pages stems from the experience the Met Office has as an operational weather prediction center. The Area Forecast Verification System \citep[AFVS;][]{sharpe13}, which was originally developed to verify weather prediction products such as the shipping forecast, has been used to verify MOSWOC products with geomagnetic storm and solar flare forecasts currently being analysed routinely. Note the Met Office shipping forecast (\hbox{http://www.metoffice.gov.uk/public/weather/marine-shipping-forecast}) provides mariners with warnings and forecasts of the conditions of the waters surrounding the British Isles.

\begin{figure}
\centering
\noindent\includegraphics[width=\textwidth]{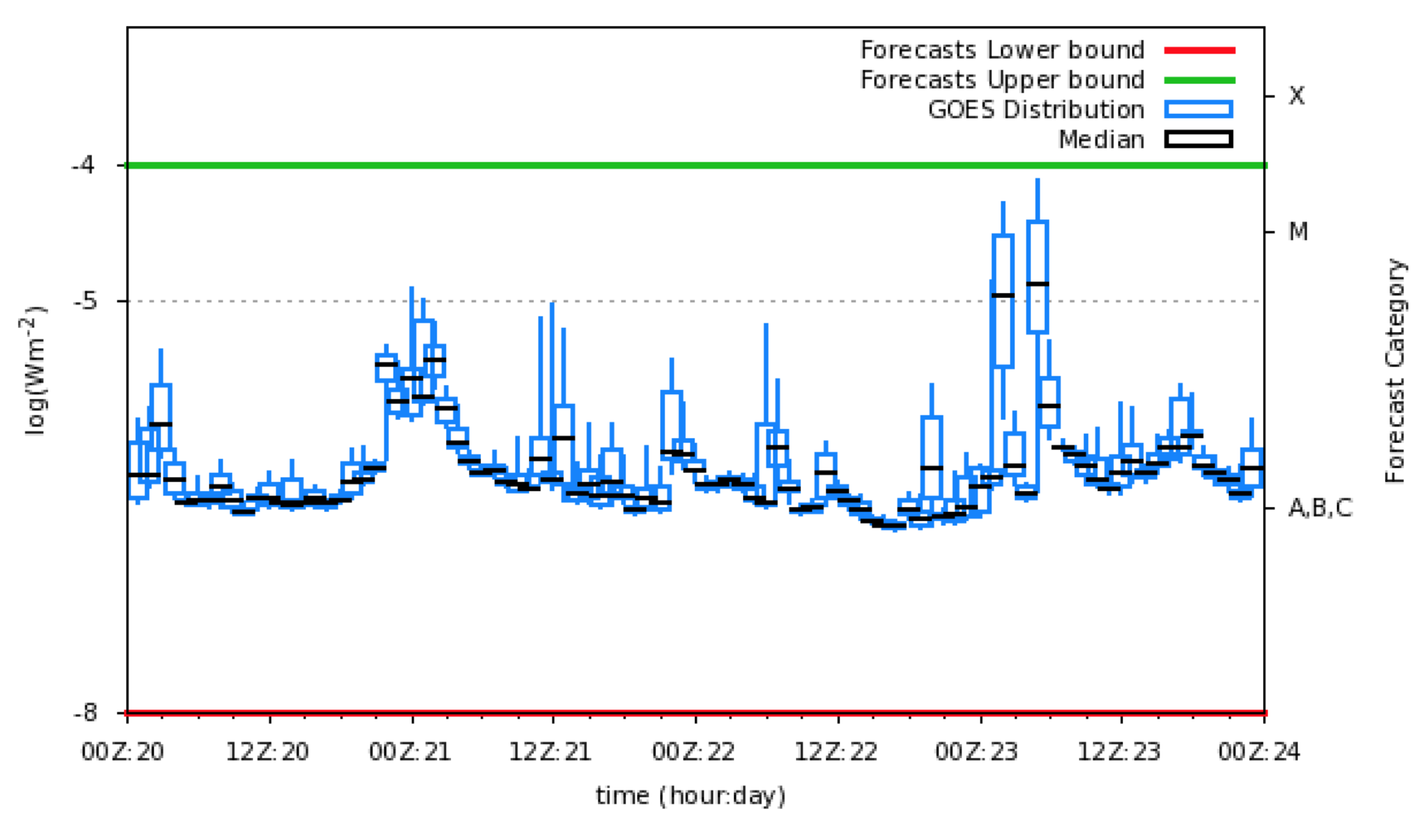}
\caption{Box-plot-line graph comparing the 2016 July 20 00:00Z forecast and X-ray flux measurements taken by the GOES-15 satellite. The x-axis displays the total range of the four day forecast (time displayed in hour:day format) with green and red horizontal lines denoting the maximum and minimum flux values to be predicted with a probability of 1\% or more (the minimum allowable probability). Box-and-whisker plots display the hourly range of observed flux values (reported every minute by GOES-15) during each hour of the forecast; the box denotes the inter-quartile range, the whiskers show the minimum and maximum observed values during each hour, and the small horizontal black line is the median observed flux value during this hour.}
\label{tot_boxplot}
\end{figure}

\begin{figure}
\centering
\noindent\includegraphics[width=\textwidth]{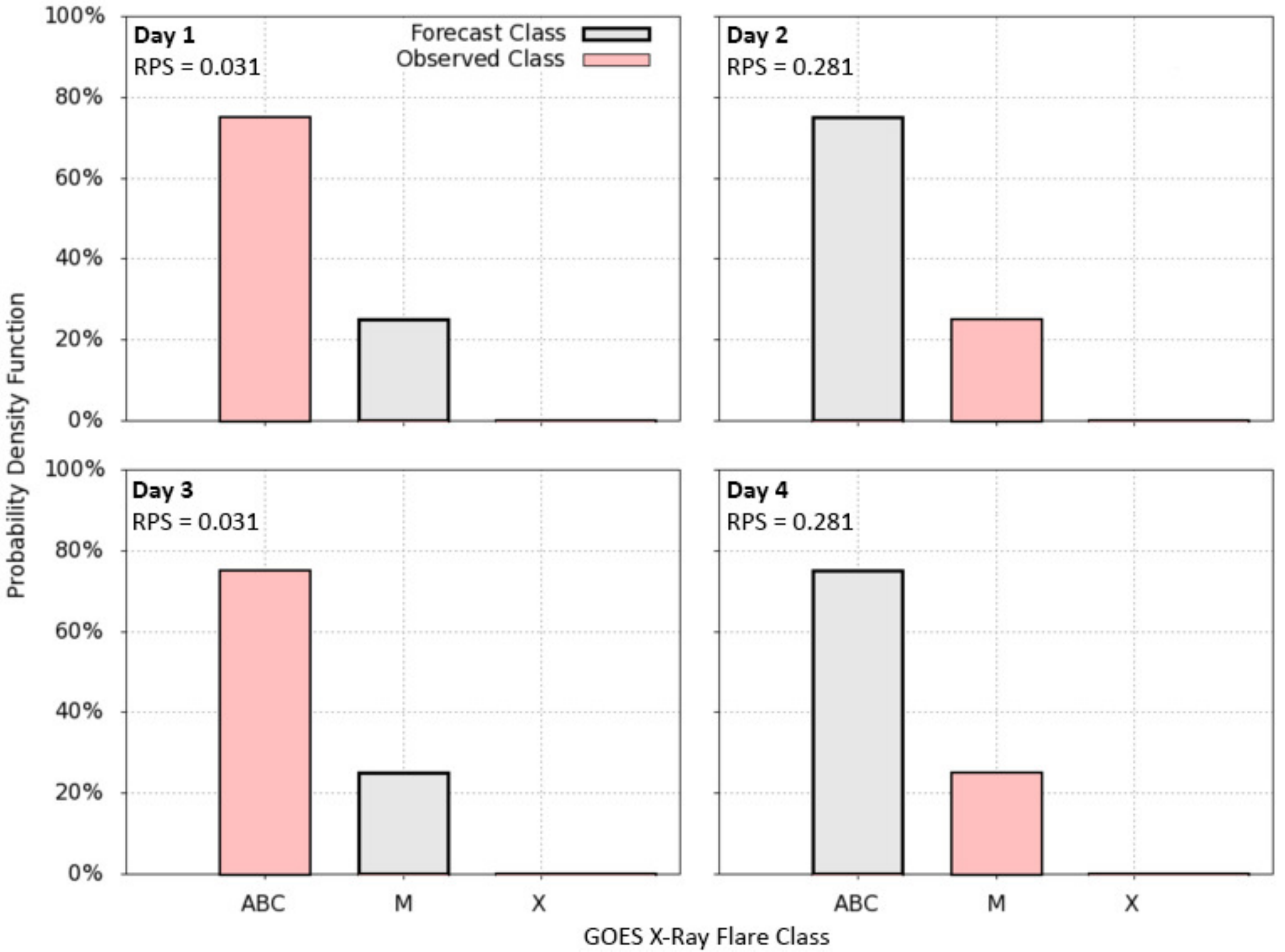}
\caption{Histograms displaying each day of the MOSWOC forecast issued for 2016 July 20 00:00Z, the observed class (pink), and the Ranked Probability Score. Each histogram refers to a different 24 h period - Day 1 refers to 00Z:20 to 00Z:21, day 2 refers to 00Z:21 to 00Z:22, Day 3 refers to 00Z:22 to 00Z:23, and Day 4 refers to 00Z:23 to 00Z:24. The x-axis displays the flux class categories and the y-axis the forecast probabilities; a pink bar denotes the maximum category observed that day and a grey bar indicates a category that was not the maximum to occur that day. The Ranked Probability Score for each day of the forecast is displayed in the top left hand corner of each plot. In this particular example the maximum category to occur was predicted to be M-class with a probability of 25\% or A/B/C-class with a probability of 75\%, and these probabilities were the same on every day of the forecast.}
\label{tot_rps}
\end{figure}

The Met Office verification pages currently focus on methods used by the terrestrial weather verification system, such as the Ranked Probability Score \citep[RPS;][]{epstein69,murphy69,murphy71}. RPS shows how well the probability forecast predicts the category that the observation fell into (in this case the GOES flare class). The RPS is calculated for recent as well as archived forecasts on the internal verification web pages and a rolling 12-month performance plot is updated daily to monitor the rolling skill. Figures~\ref{tot_boxplot} and \ref{tot_rps} show an example of the output generated by the AFVS in near-real-time for the 2016 July 20 midnight forecast. Note that `Z' in both of the Figures (format hhZ:dd) refers to Zulu military time, commonly used in operational weather prediction, and Universal Time is the same for this case.

The observed GOES long wave (1 - 8 \AA~passband) X-ray flux is displayed in Figure~\ref{tot_boxplot}. The left hand y-axis measures the X-ray flux using a $log_{10}$ scale in $Wm^{-2}$ and the right-hand y-axis displays the corresponding flare class category. On the right-hand y-axis any category below M-class is denoted by A, B, C; the lower boundary of A-class is $10^{-8} Wm^{-2}$ and the lower boundary of M-class is $10^{-5} Wm^{-2}$, consequently the X-ray flux range for category A, B, C is [$1 \times 10^{-8}$, $1 \times 10^{-5}$]. Similarly the flux range for category M is [$1 \times 10^{-5}$, $1 \times 10^{-4}$] and for category X it is $1 \times 10^{-4}$ or more. In each case the category label (A, B, C; M and X) is placed in the middle of the range on the right hand axis and each category boundary ($10^{-8} Wm^{-2}$, $10^{-5} Wm^{-2}$ and $10^{-4} Wm^{-2}$) is displayed as a dotted horizontal line. For completeness the full forecast range is included within the red and green horizontal lines, however Figure~\ref{tot_boxplot} is always shown in conjunction with the X-ray flare forecast table (as displayed on the right hand side of Figure~\ref{method:guidance_example}) so that it is easy to obtain extra information. Figure~\ref{tot_boxplot} is updated automatically every hour, providing an instant comparison between GOES X-ray flare values and the issued forecast, thereby giving busy forecasters the ability to quickly  compare every forecast against all the available observations in real-time on a single web-page. Since the data is stored (almost) indefinitely this system also facilitates easy, instant, post-event analysis which, when time allows,  gives forecasters an easy way to review their performance and (potentially) learn from any mistakes.  

Although the forecast is attempting to predict all the flux categories that are observed during each day, the AFVS assesses how accurately it predicts the strongest category to occur. The results of this analysis are shown in Figure~\ref{tot_rps} for each day of the forecast displayed in Figure~\ref{tot_boxplot}. The histogram displays the forecast percentage probabilities, with pink denoting the maximum category that was observed. During each day there was a 25\% probability that the maximum flare class was M, and a 75\% probability that the maximum flare class was either A, B, or C. Figure~\ref{tot_boxplot} indicates that C was the maximum class to occur on days one and three so the A/B/C category is shaded pink in these figures and the RPS (being a negatively orientated score) has a small value of 0.031, whereas on days two and four M was the maximum class to be observed, and the RPS has a larger value of 0.281. The WWRP/WGNE group gives a full description of how RPS is calculated (\hbox{http://www.cawcr.gov.au/projects/verification/}).

To enable a near-real-time analysis, GOES-15 X-ray flux values are obtained automatically (via ftp-download from the website maintained by SWPC) as soon as they are available. The AFVS runs hourly, generating an up-to-date analysis and producing data for on-demand plotting by MOSWOC forecasters, assisting them to issue updates as required. A histogram for each day of the forecast is displayed as soon as GOES data is available; therefore, in the example shown in Figure~\ref{tot_rps}, a Day 1 plot is available as soon as data is available for the 20th July and this plot is updated hourly until 00Z 21st July when all the data for the previous day is available, at which point the first Day 2 plot is generated and updated hourly until 00Z on 22nd July, and so on to Day 4. The forecaster, on viewing the histogram (and associated RPS), together with his/her forecast table and the box-plot-line-graph is able to instantly assess whether the original forecast is sufficiently accurate or whether the X-ray flux measurements are departing from those that were anticipated when the original forecast was issued. If the original forecast is considered insufficiently accurate alterations can be made to the mid-day update, and in exceptional circumstances it is possible to issue an emergency update. In this particular example, the mid-day forecast update on 26th July contained probabilities of 70\%, 28\% and 2\% for A/B/C, M and X class flares respectively. These small changes represent a slight increase in the perceived risk that the ensuing step-change in X-ray flux may occur before midnight and might be more severe than first thought; however, in this particular case, the step change occurred just after midnight and only just breached the M-class flare threshold of $1 \times 10^{-5}Wm{-2}$. 


\section{Discussion and Conclusions}
\label{discussion}

The \citet{gallagher02} method described in Section~\ref{method} generally performs fairly compared to other current flare forecasting methods \citep[see, e.g.,][]{bloomfield12,barnes16}, although it was not among the top-performing groups consisting of more complex methods, and as already mentioned no one method is a clear leader among the sparse comparisons undertaken thus far. The Poisson method is, however, a simple method for operational purposes, and provides an easily-understood basis from which the MOSWOC forecasters can issue official forecasts. The results of this work show that the human-edited forecasts issued by MOSWOC outperform this simple model output, thus clearly highlighting the important role of forecasters in operational space weather services. The merit of human forecasting has been discovered in previous studies, for example the performance of the ensemble developed by \citet{guerra15} improved after including human-adjusted probabilities from NOAA. When comparing the Max Millennium forecasts to other expert forecasting systems methods, \citet{bloomfield16} found that those forecasts involving human decisions were still toward the top of the performance table when considering various skill scores.

The human influenced forecasts do not perform as well at longer forecast lead times however, with a decreasing skill observed for the full-disk Day 2 - 4 forecasts. The forecasters have only the Day 1 model output to guide their decision on forecast probabilities, and thus the Day 2 - 4 IFF are mainly based on forecaster intuition. Significantly different issued probabilities to the Day 1 percentage probability are only generally seen if large active regions are due to leave or return to the solar disk. Improved backside monitoring may be useful here, such as the future L5 Carrington mission \citep{trichas15,mackay16}, however it is unlikely this would impact the overall tendency to over-forecast. Other human-influenced forecasts also show this tendency, for example \citet{bloomfield16} found a minor level of over-prediction compared to some of the published works it compared to. The forecasts issued by SWPC (see \hbox{http://www.swpc.noaa.gov/content/solar-activity-forecast-verification}) similarly show over-forecasting with reduced skill on later days. This is unsurprising, neither are the results in Table~\ref{table:iff}, since MOSWOC forecaster training is very similar to that of SWPC. In fact, many new MOSWOC forecasters spend time at SWPC headquarters shadowing SWPC forecasters, and this relationship continues with daily telecons to discuss current space weather conditions. Whilst MOSWOC produces space weather guidance tailored to possible UK impacts, they ensure any issued guidance is consistent with that of their global partners. 

Only an example of the RPS results from the AFVS are shown in the present study, however, an ideal analysis of X-ray flare forecasts would be to also employ a reference forecast against which to assess the performance using the Ranked Probability Skill Score \citep[RPSS;][]{weigel07}. The RPSS highlights the relative improvement of the probability forecast over a reference forecast in predicting the category that the observations fall into. Careful consideration should be given to the most appropriate reference to use and such an analysis is beyond the scope of the present study; however a rolling 12-month analysis using the RPSS together with other flexible verification techniques will be described in detail in a future publication. This RPSS rolling analysis will allow further investigation into how the skill of the forecaster changes over time. Whilst verifying SWPC forecasts, \citet{crown12} found experience level had little effect on the traditional Brier Skill Score (even for those with little experience). On the other hand, \citet{devos14} found those forecasters at SIDC with a strong background in solar physics and more experience with forecasting obtained high skill scores with respect to numerical models. However the level of activity also had a substantial influence on the forecast performance, with it being typically more challenging to forecast during periods of high activity. It is certainly worth continuing this analysis in future studies to see if there is any correlation between forecast skill and MOSWOC forecaster experience.

The forecasts verified in this paper were mainly obtained from a relatively quiet period after solar maximum, with solar cycle 24 having minimal activity compared to previous cycles. Once more forecasts have been issued this work can be updated to reflect a longer forecasting period. This may allow the rarer X-class flare event forecasts to also be verified once activity picks up after solar minimum. Work is underway to investigate other verification techniques used in numerical weather prediction for this purpose \citep[see e.g.,][]{ferro11}. Collaborative work is also ongoing as part of the Horizon 2020 Flare Likelihood and Region Eruption Forecasting (FLARECAST) project, which aims to develop a  fully-automated solar flare forecasting system with unmatched accuracy with real-time verification (using the same measures described here). FLARECAST will evaluate existing predictors to identify the best performers through the use of a variety of statistical, supervised, and unsupervised techniques, and implement these best performers in a user-friendly online facility. The FLARECAST system output may prove to be a more accurate model basis for MOSWOC forecasters than the currently used Poisson-method, although it is likely that the `human influence' will be needed for flare forecasting for the foreseeable future.

\acknowledgments
S. A. Murray was partly supported by the European Union's Horizon 2020 research and innovation programme under grant agreement No. 640216 (FLARECAST project), and undertook most of the data analysis while affiliated with the Met Office. GOES-class flare event lists are made available to the community by NOAA/SWPC \\
(\hbox{http://www.swpc.noaa.gov/products/solar-and-geophysical-event-reports}). MOSWOC space weather guidance is available at the Met Office public web pages \\
(\hbox{http://www.metoffice.gov.uk/space-weather}). Current MOSWOC flare forecasts are available at the Met Office Hazard Manager \\ (http://www.metoffice.gov.uk/publicsector/hazardmanager) for eligible registered users, restricted to Category 1 and 2 organisations as defined in the UK Civil Contingencies Act 2004 (\hbox{https://en.wikipedia.org/wiki/Civil\_Contingencies\_Act\_2004}). Near-real-time full-disk issued forecasts can be viewed and historical data can be downloaded at NASA/CCMC's Flare Scoreboard (http://ccmc.gsfc.nasa.gov/challenges/flare.php). All forecast data used for this analysis is also available on GitHub (\hbox{https://github.com/sophiemurray/moswoc-flare-verification}; doi:10.5281/zenodo.344886). Note that the only forecast data needed for replication of these results are simply the probabilities and time of issue rather than the guidance documents shown here as examples of operational output. We thank the editor and anonymous referees for their useful suggestions to improve this study.

\listofchanges

\end{document}